\documentclass[twocolumn,letterpaper,aps,prl,longbibliography,superscriptaddress,showpacs,floatfix]{revtex4-1}

\usepackage{graphicx}	
\usepackage{xspace}	
\usepackage{amsmath}	

\newcommand{\sqsn}{\mbox{$\sqrt{s_{_{NN}}}$}\xspace}

\newcommand{\ppb}{\mbox{$p$$+$Pb}\xspace}
\newcommand{\dau}{\mbox{$d$$+$Au}\xspace}

\newcommand{\pau}{\mbox{$p$$+$Au}\xspace}

\newcommand{\la}{\langle}
\newcommand{\ra}{\rangle}

\newcommand{\eps}{\varepsilon}
\newcommand{\mean}[1]{\la #1 \ra}
\newcommand{\dmean}[1]{\la\la #1 \ra\ra}

\newcommand{\vnt}{v_n\{2\}}
\newcommand{\vnf}{v_n\{4\}}
\newcommand{\vns}{v_n\{6\}}
\newcommand{\cnt}{c_n\{2\}}
\newcommand{\cnf}{c_n\{4\}}
\newcommand{\ccns}{c_n\{6\}}

\newcommand{\vtt}{v_2\{2\}}
\newcommand{\vtf}{v_2\{4\}}
\newcommand{\vts}{v_2\{6\}}

\newcommand{\ctf}{c_2\{4\}}

\newcommand{\vtg}{v_2\{2,|\Delta\eta|>2\}}
\newcommand{\snn}{\sqrt{s_{_{NN}}}}
\newcommand{\nfvtxt}{N_{\rm tracks}^{\rm FVTX}}

\begin{document}

\title{Measurements of multiparticle correlations in $d$$+$Au collisions 
at 200, 62.4, 39, and 19.6~GeV and $p$$+$Au collisions at 200~GeV and 
implications for collective behavior }

\newcommand{\abilene}{Abilene Christian University, Abilene, Texas 79699, USA}
\newcommand{\augie}{Department of Physics, Augustana University, Sioux Falls, South Dakota 57197, USA}
\newcommand{\banaras}{Department of Physics, Banaras Hindu University, Varanasi 221005, India}
\newcommand{\barc}{Bhabha Atomic Research Centre, Bombay 400 085, India}
\newcommand{\baruch}{Baruch College, City University of New York, New York, New York, 10010 USA}
\newcommand{\bnlcoll}{Collider-Accelerator Department, Brookhaven National Laboratory, Upton, New York 11973-5000, USA}
\newcommand{\bnlphys}{Physics Department, Brookhaven National Laboratory, Upton, New York 11973-5000, USA}
\newcommand{\caucr}{University of California-Riverside, Riverside, California 92521, USA}
\newcommand{\charlesczech}{Charles University, Ovocn\'{y} trh 5, Praha 1, 116 36, Prague, Czech Republic}
\newcommand{\chonbuk}{Chonbuk National University, Jeonju, 561-756, Korea}
\newcommand{\cns}{Center for Nuclear Study, Graduate School of Science, University of Tokyo, 6-3-1 Hongo, Bunkyo, Tokyo 113-0033, Japan}
\newcommand{\colorado}{University of Colorado, Boulder, Colorado 80309, USA}
\newcommand{\columbia}{Columbia University, New York, New York 10027 and Nevis Laboratories, Irvington, New York 10533, USA}
\newcommand{\czechtech}{Czech Technical University, Zikova 4, 166 36 Prague 6, Czech Republic}
\newcommand{\debrecen}{Debrecen University, H-4010 Debrecen, Egyetem t{\'e}r 1, Hungary}
\newcommand{\elte}{ELTE, E{\"o}tv{\"o}s Lor{\'a}nd University, H-1117 Budapest, P{\'a}zm{\'a}ny P.~s.~1/A, Hungary}
\newcommand{\eszterhazy}{Eszterh\'azy K\'aroly University, K\'aroly R\'obert Campus, H-3200 Gy\"ngy\"os, M\'atrai \'ut 36, Hungary}
\newcommand{\ewha}{Ewha Womans University, Seoul 120-750, Korea}
\newcommand{\fsu}{Florida State University, Tallahassee, Florida 32306, USA}
\newcommand{\gsu}{Georgia State University, Atlanta, Georgia 30303, USA}
\newcommand{\hiroshima}{Hiroshima University, Kagamiyama, Higashi-Hiroshima 739-8526, Japan}
\newcommand{\howard}{Department of Physics and Astronomy, Howard University, Washington, DC 20059, USA}
\newcommand{\ihepprot}{IHEP Protvino, State Research Center of Russian Federation, Institute for High Energy Physics, Protvino, 142281, Russia}
\newcommand{\illuiuc}{University of Illinois at Urbana-Champaign, Urbana, Illinois 61801, USA}
\newcommand{\inrras}{Institute for Nuclear Research of the Russian Academy of Sciences, prospekt 60-letiya Oktyabrya 7a, Moscow 117312, Russia}
\newcommand{\instpasczech}{Institute of Physics, Academy of Sciences of the Czech Republic, Na Slovance 2, 182 21 Prague 8, Czech Republic}
\newcommand{\isu}{Iowa State University, Ames, Iowa 50011, USA}
\newcommand{\jaea}{Advanced Science Research Center, Japan Atomic Energy Agency, 2-4 Shirakata Shirane, Tokai-mura, Naka-gun, Ibaraki-ken 319-1195, Japan}
\newcommand{\jyvaskyla}{Helsinki Institute of Physics and University of Jyv{\"a}skyl{\"a}, P.O.Box 35, FI-40014 Jyv{\"a}skyl{\"a}, Finland}
\newcommand{\kek}{KEK, High Energy Accelerator Research Organization, Tsukuba, Ibaraki 305-0801, Japan}
\newcommand{\korea}{Korea University, Seoul, 136-701, Korea}
\newcommand{\kurchatov}{National Research Center ``Kurchatov Institute", Moscow, 123098 Russia}
\newcommand{\kyoto}{Kyoto University, Kyoto 606-8502, Japan}
\newcommand{\lawllnl}{Lawrence Livermore National Laboratory, Livermore, California 94550, USA}
\newcommand{\losalamos}{Los Alamos National Laboratory, Los Alamos, New Mexico 87545, USA}
\newcommand{\lund}{Department of Physics, Lund University, Box 118, SE-221 00 Lund, Sweden}
\newcommand{\lyon}{IPNL, CNRS/IN2P3, Univ Lyon, Université Lyon 1, F-69622, Villeurbanne, France}
\newcommand{\maryland}{University of Maryland, College Park, Maryland 20742, USA}
\newcommand{\mass}{Department of Physics, University of Massachusetts, Amherst, Massachusetts 01003-9337, USA}
\newcommand{\michigan}{Department of Physics, University of Michigan, Ann Arbor, Michigan 48109-1040, USA}
\newcommand{\muhlenberg}{Muhlenberg College, Allentown, Pennsylvania 18104-5586, USA}
\newcommand{\nara}{Nara Women's University, Kita-uoya Nishi-machi Nara 630-8506, Japan}
\newcommand{\natmephi}{National Research Nuclear University, MEPhI, Moscow Engineering Physics Institute, Moscow, 115409, Russia}
\newcommand{\newmex}{University of New Mexico, Albuquerque, New Mexico 87131, USA}
\newcommand{\nmsu}{New Mexico State University, Las Cruces, New Mexico 88003, USA}
\newcommand{\ohio}{Department of Physics and Astronomy, Ohio University, Athens, Ohio 45701, USA}
\newcommand{\ornl}{Oak Ridge National Laboratory, Oak Ridge, Tennessee 37831, USA}
\newcommand{\orsay}{IPN-Orsay, Univ.~Paris-Sud, CNRS/IN2P3, Universit\'e Paris-Saclay, BP1, F-91406, Orsay, France}
\newcommand{\peking}{Peking University, Beijing 100871, People's Republic of China}
\newcommand{\pnpi}{PNPI, Petersburg Nuclear Physics Institute, Gatchina, Leningrad region, 188300, Russia}
\newcommand{\riken}{RIKEN Nishina Center for Accelerator-Based Science, Wako, Saitama 351-0198, Japan}
\newcommand{\rikjrbrc}{RIKEN BNL Research Center, Brookhaven National Laboratory, Upton, New York 11973-5000, USA}
\newcommand{\rikkyo}{Physics Department, Rikkyo University, 3-34-1 Nishi-Ikebukuro, Toshima, Tokyo 171-8501, Japan}
\newcommand{\saispbstu}{Saint Petersburg State Polytechnic University, St.~Petersburg, 195251 Russia}
\newcommand{\seoulnat}{Department of Physics and Astronomy, Seoul National University, Seoul 151-742, Korea}
\newcommand{\stonybrkc}{Chemistry Department, Stony Brook University, SUNY, Stony Brook, New York 11794-3400, USA}
\newcommand{\stonycrkp}{Department of Physics and Astronomy, Stony Brook University, SUNY, Stony Brook, New York 11794-3800, USA}
\newcommand{\tenn}{University of Tennessee, Knoxville, Tennessee 37996, USA}
\newcommand{\titech}{Department of Physics, Tokyo Institute of Technology, Oh-okayama, Meguro, Tokyo 152-8551, Japan}
\newcommand{\tsukuba}{Center for Integrated Research in Fundamental Science and Engineering, University of Tsukuba, Tsukuba, Ibaraki 305, Japan}
\newcommand{\vandy}{Vanderbilt University, Nashville, Tennessee 37235, USA}
\newcommand{\weizmann}{Weizmann Institute, Rehovot 76100, Israel}
\newcommand{\wigner}{Institute for Particle and Nuclear Physics, Wigner Research Centre for Physics, Hungarian Academy of Sciences (Wigner RCP, RMKI) H-1525 Budapest 114, POBox 49, Budapest, Hungary}
\newcommand{\yonsei}{Yonsei University, IPAP, Seoul 120-749, Korea}
\newcommand{\zagreb}{Department of Physics, Faculty of Science, University of Zagreb, Bijeni\v{c}ka c.~32 HR-10002 Zagreb, Croatia}
\affiliation{\abilene}
\affiliation{\augie}
\affiliation{\banaras}
\affiliation{\barc}
\affiliation{\baruch}
\affiliation{\bnlcoll}
\affiliation{\bnlphys}
\affiliation{\caucr}
\affiliation{\charlesczech}
\affiliation{\chonbuk}
\affiliation{\cns}
\affiliation{\colorado}
\affiliation{\columbia}
\affiliation{\czechtech}
\affiliation{\debrecen}
\affiliation{\elte}
\affiliation{\eszterhazy}
\affiliation{\ewha}
\affiliation{\fsu}
\affiliation{\gsu}
\affiliation{\hiroshima}
\affiliation{\howard}
\affiliation{\ihepprot}
\affiliation{\illuiuc}
\affiliation{\inrras}
\affiliation{\instpasczech}
\affiliation{\isu}
\affiliation{\jaea}
\affiliation{\jyvaskyla}
\affiliation{\kek}
\affiliation{\korea}
\affiliation{\kurchatov}
\affiliation{\kyoto}
\affiliation{\lawllnl}
\affiliation{\losalamos}
\affiliation{\lund}
\affiliation{\lyon}
\affiliation{\maryland}
\affiliation{\mass}
\affiliation{\michigan}
\affiliation{\muhlenberg}
\affiliation{\nara}
\affiliation{\natmephi}
\affiliation{\newmex}
\affiliation{\nmsu}
\affiliation{\ohio}
\affiliation{\ornl}
\affiliation{\orsay}
\affiliation{\peking}
\affiliation{\pnpi}
\affiliation{\riken}
\affiliation{\rikjrbrc}
\affiliation{\rikkyo}
\affiliation{\saispbstu}
\affiliation{\seoulnat}
\affiliation{\stonybrkc}
\affiliation{\stonycrkp}
\affiliation{\tenn}
\affiliation{\titech}
\affiliation{\tsukuba}
\affiliation{\vandy}
\affiliation{\weizmann}
\affiliation{\wigner}
\affiliation{\yonsei}
\affiliation{\zagreb}
\author{C.~Aidala} \affiliation{\michigan} 
\author{Y.~Akiba} \email[PHENIX Spokesperson: ]{akiba@rcf.rhic.bnl.gov} \affiliation{\riken} \affiliation{\rikjrbrc} 
\author{M.~Alfred} \affiliation{\howard} 
\author{V.~Andrieux} \affiliation{\michigan} 
\author{K.~Aoki} \affiliation{\kek} 
\author{N.~Apadula} \affiliation{\isu} 
\author{H.~Asano} \affiliation{\kyoto} \affiliation{\riken} 
\author{C.~Ayuso} \affiliation{\michigan} 
\author{B.~Azmoun} \affiliation{\bnlphys} 
\author{V.~Babintsev} \affiliation{\ihepprot} 
% ...
\collaboration{PHENIX Collaboration} \noaffiliation

\date{\today}


\begin{abstract}

Recently, multiparticle-correlation measurements of relativistic
$p/d/^3$He$+$Au, $p$$+$Pb, and even $p$$+$$p$ collisions have shown
surprising collective signatures.  Here we present beam-energy-scan
measurements of 2-, 4-, and 6-particle angular correlations in $d$$+$Au
collisions at $\sqrt{s_{_{NN}}}$~=~200, 62.4, 39, and 19.6~GeV.  We also
present measurements of 2- and 4-particle angular correlations in $p$$+$Au
collisions at $\sqrt{s_{_{NN}}}$~=~200~GeV.  We find the 4-particle cumulant
to be real-valued for $d$$+$Au collisions at all four energies.  We also find
that the 4-particle cumulant in $p$$+$Au has the opposite sign as that in
$d$$+$Au.  Further we find that the 6-particle cumulant agrees
with the 4-particle cumulant in $d$$+$Au collisions at 200~GeV,
indicating that nonflow effects are subdominant.
These observations provide strong evidence that the correlations
originate from the initial geometric configuration which is then
translated into the momentum distribution for all particles,
commonly referred to as collectivity.

\end{abstract}


\maketitle



One of the key discoveries at the Relativistic Heavy Ion Collider (RHIC) 
is the identification of the quark-gluon plasma (QGP) and its 
characterization as a near perfect fluid via its collective 
flow~\cite{Adcox:2004mh,Adams:2005dq,Back:2004je,Arsene:2004fa}. It has 
previously been assumed that only nucleus-on-nucleus collisions create a 
system large enough and hot enough to create the QGP.  However, five 
years ago, collective signatures were discovered in \ppb collisions at 
$\snn$~=~5.02~TeV at the large hadron collider 
(LHC)~\cite{CMS:2012qk,Abelev:2012ola,Aad:2012gla}.  Since then, similar 
evidence has been observed in $p/d/^3$He$+$Au collisions at 
$\snn$~=~200~GeV at 
RHIC~\cite{Adare:2013piz,Adare:2014keg,Adare:2015ctn,Aidala:2016vgl} and 
high-multiplicity $p$$+$$p$ collisions at \sqsn~=~2.76--13 TeV at the 
LHC~\cite{Khachatryan:2010gv,Aad:2015gqa,Khachatryan:2016txc}. 
Additionally, collective signatures at the LHC have been found not only 
with 2-particle correlations, but with multiparticle correlations as 
well~\cite{Aad:2013fja,Chatrchyan:2013nka,Abelev:2014mda,Khachatryan:2015waa}. 
Multiparticle correlations are not a unique signature of a 
hydrodynamically flowing medium~\cite{Loizides:2016tew,Dusling:2017aot}, 
and thus it is imperative that all calculational frameworks make 
quantitative predictions for these correlations. This Letter presents 
the measurement of multiparticle correlations in \dau collisions as part 
of a beam energy scan at $\snn$~=~200, 62.4, 39, and 19.6~GeV, as well 
as in \pau collisions at $\snn$~=~200~GeV.


The azimuthal distribution of particles produced in a collision can be 
described by a Fourier series with harmonic coefficients $v_n$ where $n$ 
is the harmonic number~\cite{Voloshin:1994mz}. This analysis uses direct 
calculations of cumulants~\cite{Bilandzic:2010jr}. The 2-particle 
correlator is

\begin{equation}
\mean{2} = \mean{\cos(n(\phi_1-\phi_2))} = \mean{v_n^2},
\end{equation}
where $\phi_{1,2}$ denote the azimuthal angles of two different particles
in a single event
and the single brackets denote an average over particles in a single event.
The 4-particle correlator is
\begin{equation}
\mean{4} = \mean{\cos(n(\phi_1+\phi_2-\phi_3-\phi_4))} = \mean{v_n^4},
\end{equation}
where $\phi_{1,2,3,4}$ denote the azimuthal angles of four different particles
in a single event. Finally, the 6-particle correlator is
\begin{equation}
\mean{6} = \mean{\cos(n(\phi_1+\phi_2+\phi_3-\phi_4-\phi_5-\phi_6))} = \mean{v_n^6},
\end{equation}
where $\phi_{1,2,3,4,5,6}$ denote the azimuthal angles of six different 
particles in a single event. Quite generally, any $m$-particle 
correlation will have contributions from lower-order correlations, and 
$m$-particle cumulants $c_n\{m\}$ are constructed to remove these.  In 
the case of the 2-particle cumulant, the relation is simply
\begin{equation}
\cnt = \dmean{2}, \label{eqn:cnt}
\end{equation}
where the double bracket indicates first an average over particles in a 
single event and then an average over events.  In the case of the 4- and 
6-particle cumulant, the relations are
\begin{align}
\cnf &= \dmean{4} - 2\dmean{2}^2  \quad \textrm{and} \label{eqn:cnf}\\
\ccns &= \dmean{6} - 9\dmean{4}\dmean{2} + 12\dmean{2}^3 \label{eqn:cns},
\end{align}
where it can be seen by construction that the lower-order correlations 
are removed. The harmonic coefficients are related to the cumulants by
\begin{align}
\vnt &= (\cnt)^{1/2}, \label{eqn:vnt} \\
\vnf &= (-\cnf)^{1/4} \quad \textrm{and} \label{eqn:vnf} \\
\vns &= \left(\frac{1}{4}\ccns\right)^{1/6}. \label{eqn:vns}
\end{align}

In this Letter we focus on the second harmonic, $n=$~2, which is 
interpreted as arising from elliptic flow.  For a given event category, 
there can be event-by-event differences in the strength of the elliptic 
flow.  In this case the observed $v_2$ is not a single value but rather 
a distribution.  The different cumulants have different sensitivities to 
the fluctuations of the $v_2$ distribution.  The $\vtt$ has a positive 
contribution from the variance of the distribution, whereas $\vtf$ and 
$\vts$ have negative contributions from the variance.  Comparisons of 
the different cumulants can yield insights into not only the central 
value of the $v_2$ but also the nature of its event-by-event 
fluctuations.

Not all angular correlations are global in nature.  The term nonflow is 
used to describe angular correlations arising from anything not 
considered global or collective in nature, and typically includes 
resonance decays, quantum interference correlations, Coulomb 
interactions, jet correlations, etc. Most of these generate correlations 
among only a small subset of the total produced particles, thus 
4-particle correlations are typically much less sensitive than 
2-particle correlations to nonflow effects. For that reason, comparison 
between 2-, 4-, and 6-particle correlations can also yield insights into 
nonflow effects.  Considering the event-by-event $v_2$ fluctuations (in 
the Gaussian limit) and nonflow, one has

\begin{align}
\label{eqn:v22sigmadelta} \vtt &= (v_2^2 + \sigma^2 + \delta^2)^{1/2} \quad \textrm{and} \\
\label{eqn:v24sigma} \vtf \approx \vts &\approx (v_2^2 - \sigma^2)^{1/2},
\end{align}
where $\sigma^2$ is the variance of the distribution and $\delta^2$ 
parameterizes the nonflow~\cite{Ollitrault:2009ie}.


In 2016, the PHENIX experiment~\cite{Adcox:2003zm} at RHIC collected 
data from \dau collisions at four different energies ($\snn$~=~200, 
62.4, 39, and 19.6~GeV).  In 2015, data from \pau collisions at 
$\snn$~=~200~GeV was collected.  PHENIX triggered on minimum bias  
and high multiplicity events utilizing a beam beam counter 
(BBC)~\cite{Ikematsu:1998fm} at 200 and 62.4~GeV or a forward silicon 
detector (FVTX)~\cite{Aidala:2013vna} at 39 and 19.6~GeV. Using 
information from the BBC and FVTX, we require events to have a collision 
vertex within $|z|<$ 10~cm of the nominal center of the PHENIX 
coordinate system.

The particle correlations are formed from reconstructed tracks in the 
FVTX, which has two arms covering $-3<\eta<-1$ and $+1<\eta<+3$ in 
pseudorapidity. The FVTX does not provide momentum information, but 
simulations have determined that the efficiency is momentum independent 
for $p_T \gtrsim$ 0.3~GeV/$c$. We require tracks in the FVTX to have a 
distance of closest approach (DCA) to the reconstructed vertex less than 
2~cm and to have hits in at least 3 of the 4 layers of the FVTX. We 
evaluate all quantities as a function of the number of reconstructed 
tracks in the FVTX, $\nfvtxt$.  The $\dmean{6}$, $\dmean{4}$, and 
$\dmean{2}$ are evaluated in events categorized by a single integer 
value of $\nfvtxt$.  Event categories are then combined into wider bins 
as needed to achieve adequate statistical precision. As an illustrative 
example, 10$<\nfvtxt<$30 corresponds to
centralities
in \dau of
1.3\%--52\%,
4.1$\times$10$^{-2}$\%--33\%,
6.5$\times$10$^{-4}$\%--21\%,
and 3.3$\times$10$^{-6}$\%--10\%
at 200, 62.4, 39, 19.6 GeV respectively,
and in \pau at 200 GeV of
0.22\%--29\%.

\begin{figure*}[hbtp]
\includegraphics[width=0.45\linewidth]{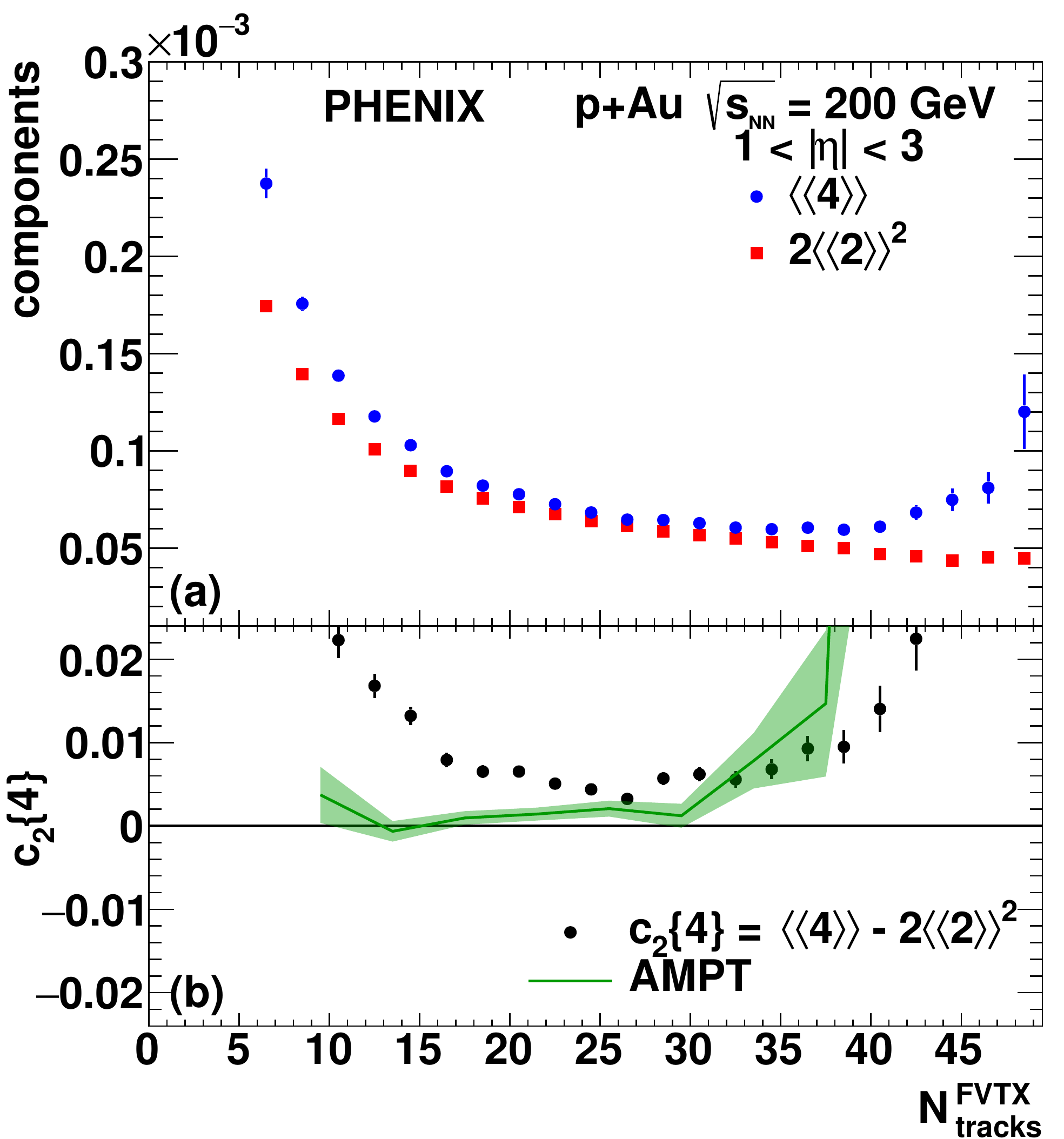}
\includegraphics[width=0.45\linewidth]{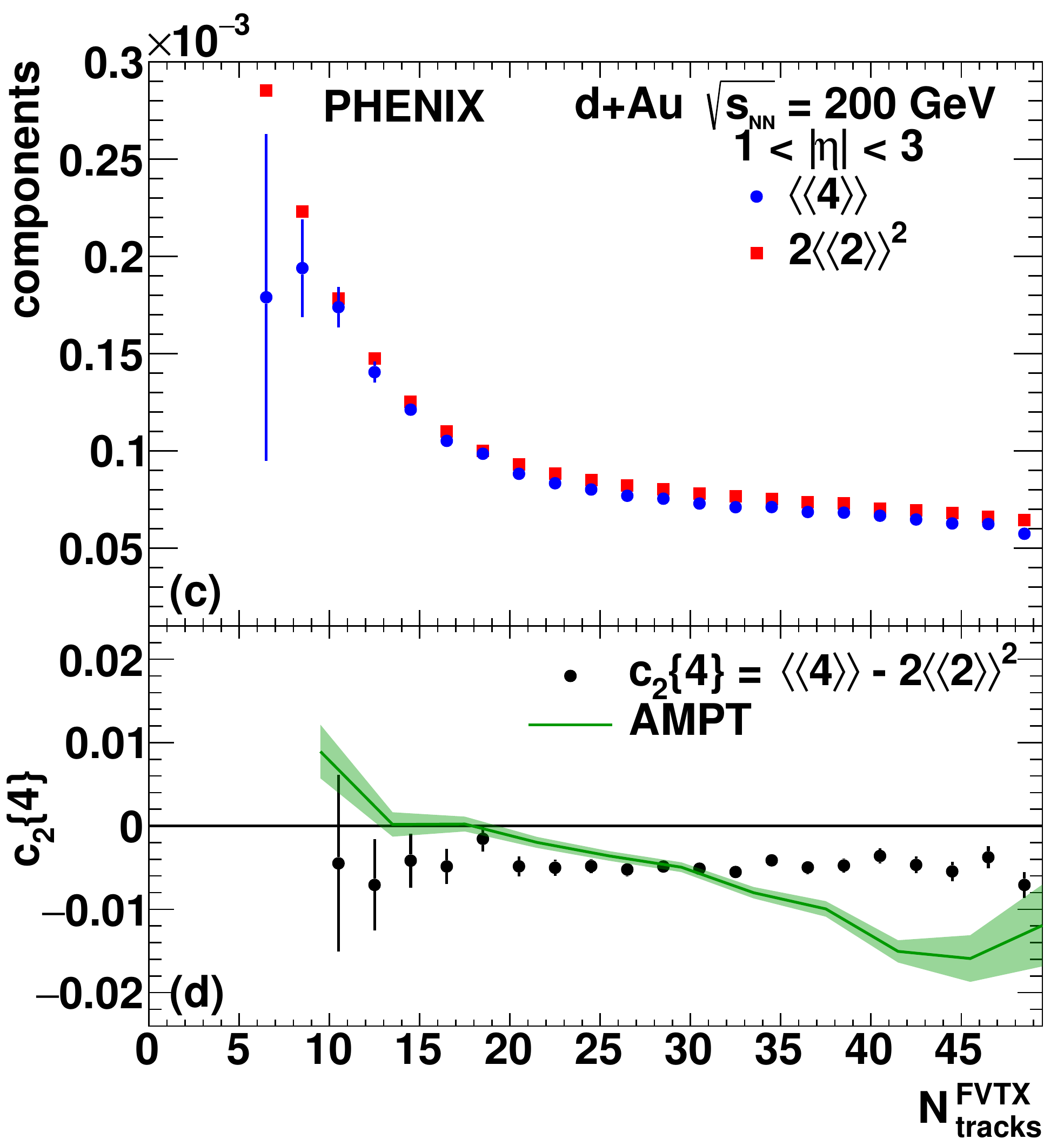}
\caption{Components $\dmean{4}$ and $2\dmean{2}^2$ and cumulant $\ctf = 
\dmean{4} - 2\dmean{2}^2$ as a function of $\nfvtxt$.  (a) and (b) 
show the components and cumulant, respectively, in \pau collisions 
at $\snn=$~200~GeV.  (c) and (d) show the components and 
cumulant, respectively, in \dau collisions at $\snn=$~200~GeV.   
(b) and (d) also show the cumulant as measured in AMPT for \pau 
and \dau, respectively, indicated by the green line.  The shaded green 
band indicates the statistical uncertainty on the AMPT values.}
\label{fig:components}
\end{figure*}

Figure~\ref{fig:components} shows (a,c) the $\dmean{4}$ and 
$2\dmean{2}^2$ and (b,c) cumulant $\ctf$ for (a,b) \pau collisions and 
(c,d) \dau collisions at $\snn=$~200~GeV.  In both cases, only 
statistical uncertainties are shown. The cumulant in \pau is positive, 
indicating that $\vtf$ is complex. In contrast, in \ppb collisions at 
$\snn=$~5.02~TeV, the cumulant is negative and the $\vtf$ is real for 
sufficiently high 
multiplicity~\cite{Aad:2013fja,Chatrchyan:2013nka,Abelev:2014mda,Khachatryan:2015waa}. 
However, the cumulant in \dau collisions at $\snn$~=~200~GeV is 
negative, indicating that $\vtf$ is real.  For now, we focus on the \dau 
results and will return to the \pau system later.

\begin{figure*}[hbtp]
\includegraphics[width=0.99\linewidth]{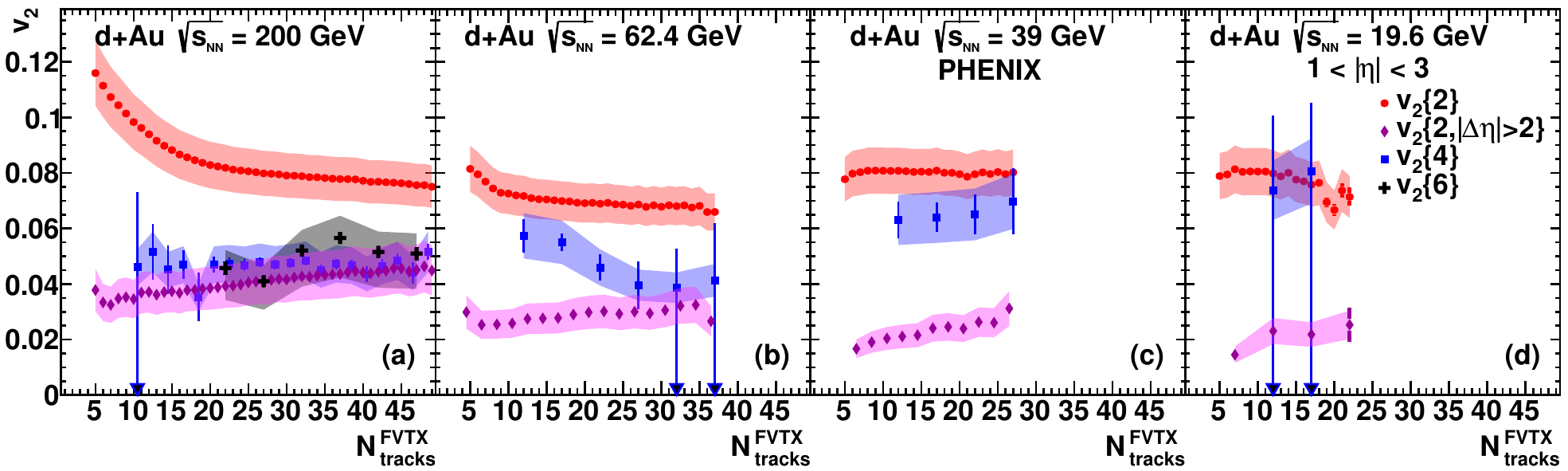}
\caption{$\vtt$, $\vtg$, and $\vtf$ as a function of $\nfvtxt$ in \dau 
collisions with $\snn$ = (a) 200~GeV, (b) 62.4~GeV, 
(c) 39~GeV, and (d) 19.6~GeV; also shown in (a) is $\vts$ 
for $\snn$~=~200~GeV.
The arrowheads on the statistical uncertainties indicate cases where the
standard 1$\sigma$ uncertainty on the $\ctf$ crosses zero.  For 19.6 GeV,
the combined confidence interval for $\vtf$ to be real is 79\%.}
\label{fig:v24_4panel}
\end{figure*}

Figure~\ref{fig:v24_4panel} shows the calculated $\vtt$ and $\vtf$ in 
\dau collisions at 200, 62.4, 39, and 19.6~GeV. Systematic 
uncertainties, shown as colored bands, are point-to-point correlated and 
are determined as the quadrature sum of the following contributions. We 
vary the event vertex cut from the 10~cm default to 5~cm as a check on 
the $z$ dependence of the FVTX acceptance and find a systematic 
uncertainty of approximately 1\% (10\%) for 2-particle (4-particle) 
correlations.  The DCA cut is varied from the default 2~cm cut to 
1.5~cm, and we find a systematic difference of approximately 1\%. The 
azimuthal acceptance in the FVTX is not uniform due to detector 
inefficiencies, so corrections need to be applied.  We use the Q-vector 
recentering method~\cite{Poskanzer:1998yz} as the default and compare to 
the isotropic terms in Ref.~\cite{Bilandzic:2010jr}. We assess an 
uncertainty of 10\% of the value of the $\vtt$ and $\vtf$ due to this 
correction, which is the dominant source of systematic uncertainty.

Rather strikingly, we observe real-valued $\vtf$ in \dau at all four 
collision energies. This is additional evidence in support of collective 
behavior in small 
systems~\cite{Adare:2013piz,Adare:2014keg,Adare:2015ctn,Aidala:2016vgl}.  
The same patterns seen in \ppb collisions at the LHC appears to persist 
in \dau at collision energies a factor of 250 lower.

Further, Fig.~\ref{fig:v24_4panel} shows the $\vts$ in \dau collisions 
at 200~GeV. The $\vts$ is consistent with $\vtf$ across the full 
$\nfvtxt$ range. This shows that, at least at 200~GeV, the $\vtf$ is 
dominated by flow, rather than nonflow. The statistics at the lower 
energies are not enough to determine a reliable $\vts$.

\begin{figure}[hbtp]
\includegraphics[width=1.0\linewidth]{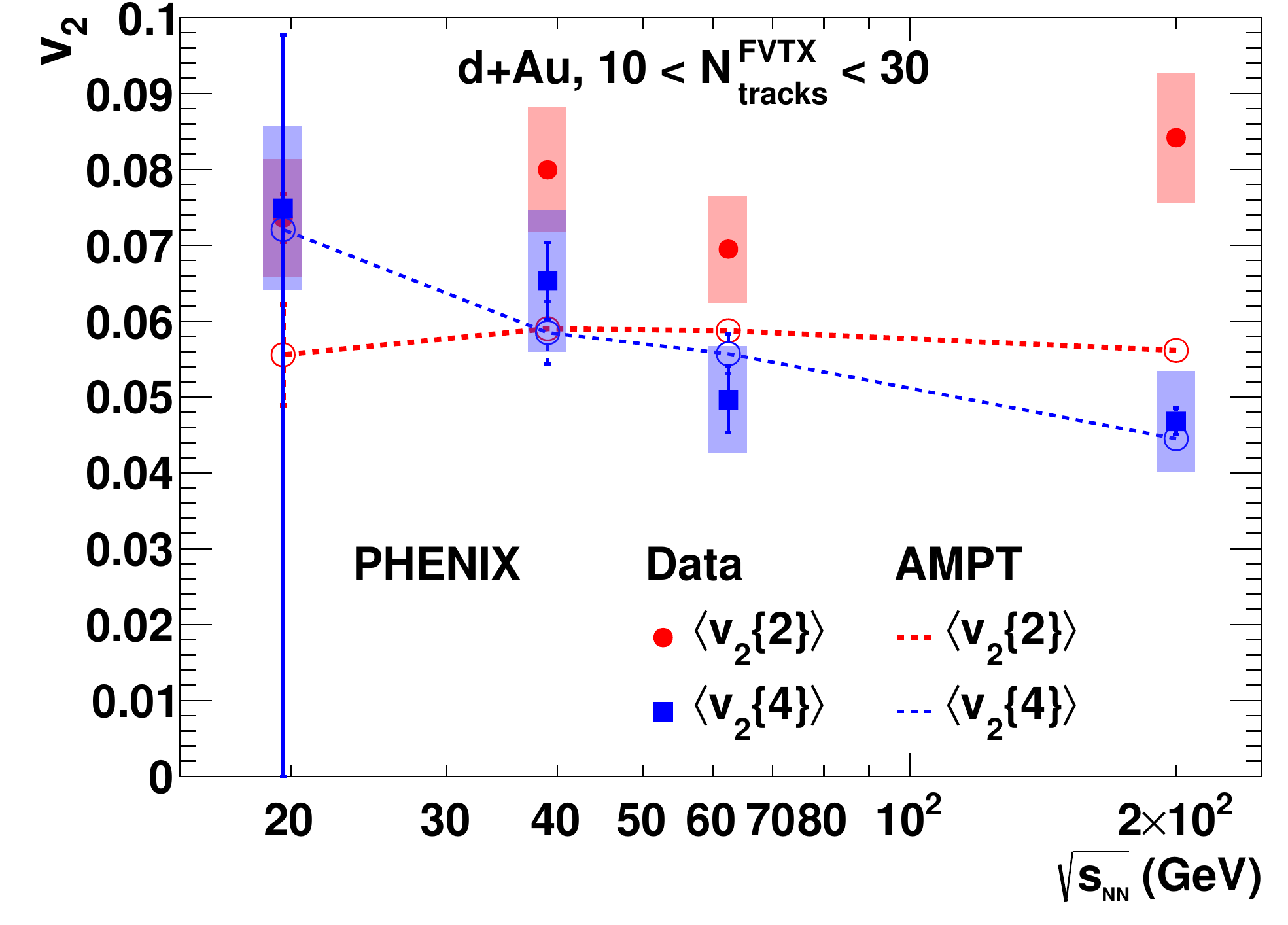}
\caption{$\vtt$ and $\vtf$ as a function of $\snn$ in \dau collisions.  
AMPT calculations are shown for comparison.
For 19.6 GeV the confidence interval for $\vtf$ to be real is 79\%.}
\label{fig:v24_avgsqrts}
\end{figure}

Figure~\ref{fig:v24_avgsqrts} shows the $\vtt$ and $\vtf$ in \dau 
collisions as a function of $\snn$ when averaged over 10~$<\nfvtxt<$~30.  
We find that $\vtf < \vtt$ at the higher energies, as expected from 
Eqns.~\ref{eqn:v22sigmadelta},~\ref{eqn:v24sigma} where both the 
event-to-event $v_2$ fluctuations and nonflow contribute positively to 
$\vtt$, and the $v_2$ fluctuations contribute negatively to $\vtf$ while 
nonflow should be significantly reduced.  However, there is a trend that 
the difference between the $\vtt$ and $\vtf$ decreases with decreasing 
energy, with $\vtt \approx \vtf$ within uncertainties at 19.6 and 39 
GeV. If Eqns.~\ref{eqn:v22sigmadelta},~\ref{eqn:v24sigma} are valid at 
these low multiplicities, the $\vtt$ and $\vtf$ may converge if the flow 
fluctuations ($\sigma$) or the nonflow ($\delta$) decrease at lower \dau 
energies. Monte Carlo Glauber calculations indicate that the 
event-by-event fluctuations in the initial geometry are quite similar 
for \dau collisions at all four energies. In the case of nonflow, while 
jet contributions decrease at lower energy, the expectation is that 
$\delta$ increases because one has a nonflow correlation from a fixed 
particle number ($N$) that is diluted by the total number of particles 
in the event ($M$), which is smaller for lower energy $d$$+$Au collisions 
even at fixed number of FVTX tracks. The measured 2-and 4-particle 
correlations appear to be more complex than the assumptions in 
Eqns.~\ref{eqn:v22sigmadelta},~\ref{eqn:v24sigma}.

To explore these trends in more detail, we utilize 
A-Multi-Phase-Transport (AMPT) model that includes parton production via 
string melting, parton scattering, hadronization via coalescence, and 
hadronic scattering~\cite{Lin:2004en}.  AMPT has been successful at 
qualitatively describing many signatures of collectivity in small and 
large collision systems~\cite{Bzdak:2014dia,Koop:2015wea,Koop:2015trj},
and we utilize the identical parameters and setup as in Ref.~\cite{Koop:2015trj}.
Modeling the FVTX acceptance and efficiency, we find reasonable agreement
with the experimental FVTX track distribution and then calculate the
$\vtt$ and $\vtf$ from AMPT as shown in Fig.~\ref{fig:v24_avgsqrts}.
The AMPT calculations include event-by-event geometry fluctuations via
Monte Carlo Glauber~\cite{Loizides:2014vua}, flow (defined here as momentum
anisotropy relative to the initial geometry), and nonflow. AMPT gives a
reasonable description of the magnitude and trend of $\vtf$, while
underpredicting the $\vtt$; this may be due to an underestimation of the
nonflow.

Our measurement of $\vtt$ is particularly susceptible to nonflow 
contributions because we allow combinations that may be close in 
pseudorapidity.  Analyses of LHC data (e.g. 
Refs~\cite{Aad:2013fja,Chatrchyan:2013nka,Abelev:2014mda,Khachatryan:2015waa}) 
introduce a pseudorapidity gap $|\Delta \eta| > 2$ between all pairs 
thus reducing contributions from particle decays, intrajet 
correlations, etc. In our case, because of the FVTX acceptance, such an 
$\eta$ gap necessitates requiring one particle per arm.  In \dau 
collisions, particularly at the lower energies, this means that the 
kinematics for the $\vtg$ and $\vtf$ are very different and the former 
will be strongly effected by asymmetries in $v_2$ between forward and 
backward rapidity, as well as longitudinal 
decorrelations~\cite{Petersen:2011fp,Xiao:2012uw}.

Nonetheless, we calculate $\vtg$ and show the results in 
Fig.~\ref{fig:v24_4panel}. We find that $\vtg<\vtt$ for all four 
energies as expected from the reduction in nonflow contributions; 
however, we also find that $\vtg<\vtf$, which cannot be reconciled 
within the context of Eqns.~\ref{eqn:v22sigmadelta},~\ref{eqn:v24sigma} 
alone.  In AMPT, the true $v_2$ at forward (d-going) rapidity $v_2^F$ is 
significantly lower than $v_2$ at backward (Au-going) rapidity $v_2^B$.  
The $\vtg~=~\sqrt{v_2^Bv_2^F}$ whereas the $\vtf$ is heavily weighted 
towards $v_2^B$ where there are more tracks in the FVTX.  This 
difference in kinematic sensitivity makes a quantitative comparison with 
$\vtf$ challenging, while opening the door to new sensitivity to the 
longitudinal structure of the correlations.

\begin{figure}[hbtp]
\includegraphics[width=1.0\linewidth]{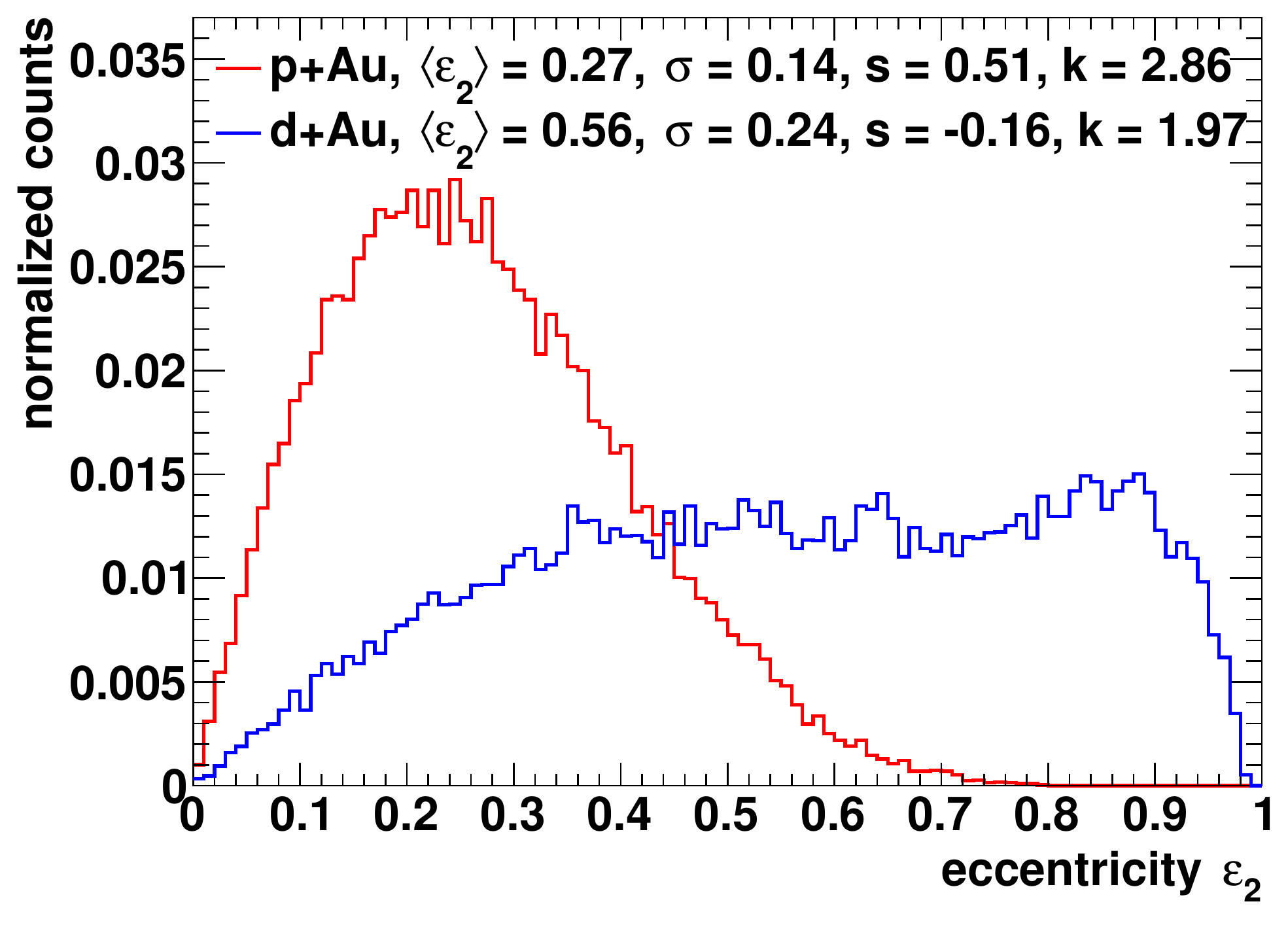}
\caption{Eccentricity distributions for \pau and \dau at 
$\snn$~=~200~GeV as calculated via Monte Carlo Glauber.  The exact 
values for the mean $\mean{\eps_2}$, standard deviation $\sigma$, 
skewness $s$, and kurtosis $k$ are listed on the figure in the caption
for each distribution.}
\label{fig:ecce}
\end{figure}

Let us now return to the results in \pau collisions, where the $\vtf$ is 
complex. Following Eqn.~\ref{eqn:v24sigma}, if the event-by-event $v_2$ 
fluctuations are larger in \pau compared with \dau to the extent that 
$\sigma > v_{2}$, this would explain the sign change.  In the case of 
ideal hydrodynamic evolution, the flow $v_2$ is proportional to the 
initial elliptical geometric eccentricity 
$\eps_2$~\cite{PhysRevD.46.229}. Thus, we show in Fig.~\ref{fig:ecce} 
the $\eps_2$ distributions from Monte Carlo Glauber 
calculations~\cite{Loizides:2014vua} for \pau and \dau at 
$\snn$~=~200~GeV. The average $\eps_2$ for \dau is almost twice the 
value for \pau, and both distributions are highly nonGaussian. The 
$\eps_2$ distribution in \pau collisions has large positive skew and the 
$\eps_2$ distribution in \dau collisions is significantly platykurtic.  
The exact values of the skewness $s$ and kurtosis $k$ are listed in the 
figure.  We can define cumulants of $\eps_2$ exactly as one does for the 
$v_2$ in Eqs.~\ref{eqn:cnt}--\ref{eqn:vns}. If we do not restrict 
ourselves to the Gaussian approximation, but instead include all higher 
moments, we find $\eps_2\{4\}$ values of 0.166 (0.508) in \pau (\dau) 
collisions when using the exact form compared to 0.232 (0.505) in the 
Gaussian approximation. The conventional Gaussian approximation 
significantly overpredicts the exact calculation in \pau, and slightly 
underpredicts it in \dau. 
These geometry fluctuation 
contributions go in the right direction to reducing the magnitude of the 
$\vtf$ in \pau collisions, but not to the extent of flipping the sign of 
the cumulant and generating a complex $\vtf$.

It is possible that fluctuations in translating the initial eccentricity 
into the final state momentum anisotropy lead to additional fluctuations 
in the $v_2$ values that could result in $\ctf$ becoming positive in 
\pau collisions. In fact, calculations utilizing AMPT, which describe 
the negative $\ctf$ and thus real $\vtf$ in \dau, yield a positive 
valued $\ctf$ in \pau collisions, as shown by the green curves in 
Fig.~\ref{fig:components}.
It is notable that these AMPT calculations utilize the identical
Monte Carlo Glauber initial conditions as shown in Fig.~\ref{fig:ecce},
and thus this sign change is definitively from additional fluctuation
effects.


In summary, we have presented measurements of $v_2$ from multiparticle 
correlations in \pau collisions at $\snn$~=~200~GeV and in \dau 
collisions at $\snn$~=~200, 62.4, 39, and 19.6~GeV.  We find real-valued 
$\vtf$ in \dau at all collision energies, providing evidence for 
collectivity in \dau at all energies.  At the highest energy in \dau, 
this evidence is further strengthened by the observation of 
$\vtf~\approx~\vts$, indicating that nonflow contributions to $\vtf$ are 
subdominant.  We find $\vtf$ is complex in \pau at $\snn$~=~200~GeV.  
The $\eps_2$ distribution in \pau is highly nonGaussian, leading to an 
$\eps_2\{4\}$ much lower than Gaussian expectations.  Additional 
fluctuations in the translation of $\eps_2$ to $v_2$ may explain the 
observation of $\vtf$ being complex in \pau.  That collision systems 
with different initial geometries (\pau and \dau) at fixed collision 
energy (200 GeV) lead to significantly different cumulants indicates a 
geometrical and therefore collective origin of the correlations.



We thank the staff of the Collider-Accelerator and Physics
Departments at Brookhaven National Laboratory and the staff of
the other PHENIX participating institutions for their vital
contributions.  We acknowledge support from the
Office of Nuclear Physics in the
Office of Science of the Department of Energy,
the National Science Foundation,
Abilene Christian University Research Council,
Research Foundation of SUNY, and
Dean of the College of Arts and Sciences, Vanderbilt University
(U.S.A),
Ministry of Education, Culture, Sports, Science, and Technology
and the Japan Society for the Promotion of Science (Japan),
Conselho Nacional de Desenvolvimento Cient\'{\i}fico e
Tecnol{\'o}gico and Funda\c c{\~a}o de Amparo {\`a} Pesquisa do
Estado de S{\~a}o Paulo (Brazil),
Natural Science Foundation of China (People's Republic of China),
Croatian Science Foundation and
Ministry of Science and Education (Croatia),
Ministry of Education, Youth and Sports (Czech Republic),
Centre National de la Recherche Scientifique, Commissariat
{\`a} l'{\'E}nergie Atomique, and Institut National de Physique
Nucl{\'e}aire et de Physique des Particules (France),
Bundesministerium f\"ur Bildung und Forschung, Deutscher
Akademischer Austausch Dienst, and Alexander von Humboldt Stiftung (Germany),
National Science Fund, OTKA, EFOP, and the Ch. Simonyi Fund (Hungary),
Department of Atomic Energy and Department of Science and Technology (India),
Israel Science Foundation (Israel),
Basic Science Research Program through NRF of the Ministry of Education (Korea),
Physics Department, Lahore University of Management Sciences (Pakistan),
Ministry of Education and Science, Russian Academy of Sciences,
Federal Agency of Atomic Energy (Russia),
VR and Wallenberg Foundation (Sweden),
the U.S. Civilian Research and Development Foundation for the
Independent States of the Former Soviet Union,
the Hungarian American Enterprise Scholarship Fund,
and the US-Israel Binational Science Foundation.




\begin{thebibliography}{35}%
\makeatletter
\providecommand \@ifxundefined [1]{%
 \@ifx{#1\undefined}
}%
\providecommand \@ifnum [1]{%
 \ifnum #1\expandafter \@firstoftwo
 \else \expandafter \@secondoftwo
 \fi
}%
\providecommand \@ifx [1]{%
 \ifx #1\expandafter \@firstoftwo
 \else \expandafter \@secondoftwo
 \fi
}%
\providecommand \natexlab [1]{#1}%
\providecommand \enquote  [1]{``#1''}%
\providecommand \bibnamefont  [1]{#1}%
\providecommand \bibfnamefont [1]{#1}%
\providecommand \citenamefont [1]{#1}%
\providecommand \href@noop [0]{\@secondoftwo}%
\providecommand \href [0]{\begingroup \@sanitize@url \@href}%
\providecommand \@href[1]{\@@startlink{#1}\@@href}%
\providecommand \@@href[1]{\endgroup#1\@@endlink}%
\providecommand \@sanitize@url [0]{\catcode `\\12\catcode `\$12\catcode
  `\&12\catcode `\#12\catcode `\^12\catcode `\_12\catcode `\%12\relax}%
\providecommand \@@startlink[1]{}%
\providecommand \@@endlink[0]{}%
\providecommand \url  [0]{\begingroup\@sanitize@url \@url }%
\providecommand \@url [1]{\endgroup\@href {#1}{\urlprefix }}%
\providecommand \urlprefix  [0]{URL }%
\providecommand \Eprint [0]{\href }%
\providecommand \doibase [0]{http://dx.doi.org/}%
\providecommand \selectlanguage [0]{\@gobble}%
\providecommand \bibinfo  [0]{\@secondoftwo}%
\providecommand \bibfield  [0]{\@secondoftwo}%
\providecommand \translation [1]{[#1]}%
\providecommand \BibitemOpen [0]{}%
\providecommand \bibitemStop [0]{}%
\providecommand \bibitemNoStop [0]{.\EOS\space}%
\providecommand \EOS [0]{\spacefactor3000\relax}%
\providecommand \BibitemShut  [1]{\csname bibitem#1\endcsname}%
\let\auto@bib@innerbib\@empty
\bibitem [{\citenamefont {Adcox}\ \emph {et~al.}(2005)\citenamefont {Adcox}
  \emph {et~al.}}]{Adcox:2004mh}%
  \BibitemOpen
  \bibfield  {author} {\bibinfo {author} {\bibfnamefont {K.}~\bibnamefont
  {Adcox}} \emph {et~al.} (\bibinfo {collaboration} {PHENIX Collaboration}),\
  }\bibfield  {title} {\enquote {\bibinfo {title} {{Formation of dense partonic
  matter in relativistic nucleus nucleus collisions at RHIC: Experimental
  evaluation by the PHENIX collaboration}},}\ }\href@noop {} {\bibfield
  {journal} {\bibinfo  {journal} {Nucl. Phys. A}\ }\textbf {\bibinfo {volume}
  {757}},\ \bibinfo {pages} {184} (\bibinfo {year} {2005})}\BibitemShut
  {NoStop}%
\bibitem [{\citenamefont {Adams}\ \emph {et~al.}(2005)\citenamefont {Adams}
  \emph {et~al.}}]{Adams:2005dq}%
  \BibitemOpen
  \bibfield  {author} {\bibinfo {author} {\bibfnamefont {J.}~\bibnamefont
  {Adams}} \emph {et~al.} (\bibinfo {collaboration} {STAR Collaboration}),\
  }\bibfield  {title} {\enquote {\bibinfo {title} {{Experimental and
  theoretical challenges in the search for the quark gluon plasma: The STAR
  collaboration's critical assessment of the evidence from RHIC collisions}},}\
  }\href@noop {} {\bibfield  {journal} {\bibinfo  {journal} {Nucl. Phys. A}\
  }\textbf {\bibinfo {volume} {757}},\ \bibinfo {pages} {102} (\bibinfo {year}
  {2005})}\BibitemShut {NoStop}%
\bibitem [{\citenamefont {Back}\ \emph {et~al.}(2005)\citenamefont {Back} \emph
  {et~al.}}]{Back:2004je}%
  \BibitemOpen
  \bibfield  {author} {\bibinfo {author} {\bibfnamefont {B.~B.}\ \bibnamefont
  {Back}} \emph {et~al.} (\bibinfo {collaboration} {PHOBOS Collaboration}),\
  }\bibfield  {title} {\enquote {\bibinfo {title} {{The PHOBOS perspective on
  discoveries at RHIC}},}\ }\href@noop {} {\bibfield  {journal} {\bibinfo
  {journal} {Nucl. Phys. A}\ }\textbf {\bibinfo {volume} {757}},\ \bibinfo
  {pages} {28} (\bibinfo {year} {2005})}\BibitemShut {NoStop}%
\bibitem [{\citenamefont {Arsene}\ \emph {et~al.}(2005)\citenamefont {Arsene}
  \emph {et~al.}}]{Arsene:2004fa}%
  \BibitemOpen
  \bibfield  {author} {\bibinfo {author} {\bibfnamefont {I.}~\bibnamefont
  {Arsene}} \emph {et~al.} (\bibinfo {collaboration} {BRAHMS Collaboration}),\
  }\bibfield  {title} {\enquote {\bibinfo {title} {{Quark gluon plasma and
  color glass condensate at RHIC? The perspective from the BRAHMS
  experiment}},}\ }\href@noop {} {\bibfield  {journal} {\bibinfo  {journal}
  {Nucl. Phys. A}\ }\textbf {\bibinfo {volume} {757}},\ \bibinfo {pages} {1}
  (\bibinfo {year} {2005})}\BibitemShut {NoStop}%
\bibitem [{\citenamefont {Chatrchyan}\ \emph
  {et~al.}(2013{\natexlab{a}})\citenamefont {Chatrchyan} \emph
  {et~al.}}]{CMS:2012qk}%
  \BibitemOpen
  \bibfield  {author} {\bibinfo {author} {\bibfnamefont {S.}~\bibnamefont
  {Chatrchyan}} \emph {et~al.} (\bibinfo {collaboration} {CMS Collaboration}),\
  }\bibfield  {title} {\enquote {\bibinfo {title} {{Observation of long-range
  near-side angular correlations in proton-lead collisions at the LHC}},}\
  }\href {\doibase 10.1016/j.physletb.2012.11.025} {\bibfield  {journal}
  {\bibinfo  {journal} {Phys. Lett. B}\ }\textbf {\bibinfo {volume} {718}},\
  \bibinfo {pages} {795} (\bibinfo {year} {2013}{\natexlab{a}})}\BibitemShut
  {NoStop}%
\bibitem [{\citenamefont {Abelev}\ \emph {et~al.}(2013)\citenamefont {Abelev}
  \emph {et~al.}}]{Abelev:2012ola}%
  \BibitemOpen
  \bibfield  {author} {\bibinfo {author} {\bibfnamefont {B.}~\bibnamefont
  {Abelev}} \emph {et~al.} (\bibinfo {collaboration} {ALICE Collaboration}),\
  }\bibfield  {title} {\enquote {\bibinfo {title} {{Long-range angular
  correlations on the near and away side in $p$-Pb collisions at
  $\sqrt{s_{NN}}=5.02$ TeV}},}\ }\href {\doibase
  10.1016/j.physletb.2013.01.012} {\bibfield  {journal} {\bibinfo  {journal}
  {Phys. Lett. B}\ }\textbf {\bibinfo {volume} {719}},\ \bibinfo {pages} {29}
  (\bibinfo {year} {2013})}\BibitemShut {NoStop}%
\bibitem [{\citenamefont {Aad}\ \emph {et~al.}(2013{\natexlab{a}})\citenamefont
  {Aad} \emph {et~al.}}]{Aad:2012gla}%
  \BibitemOpen
  \bibfield  {author} {\bibinfo {author} {\bibfnamefont {G.}~\bibnamefont
  {Aad}} \emph {et~al.} (\bibinfo {collaboration} {ATLAS Collaboration}),\
  }\bibfield  {title} {\enquote {\bibinfo {title} {{Observation of Associated
  Near-Side and Away-Side Long-Range Correlations in $\sqrt{s_{NN}}$=5.02~TeV
  Proton-Lead Collisions with the ATLAS Detector}},}\ }\href {\doibase
  10.1103/PhysRevLett.110.182302} {\bibfield  {journal} {\bibinfo  {journal}
  {Phys. Rev. Lett.}\ }\textbf {\bibinfo {volume} {110}},\ \bibinfo {pages}
  {182302} (\bibinfo {year} {2013}{\natexlab{a}})}\BibitemShut {NoStop}%
\bibitem [{\citenamefont {Adare}\ \emph {et~al.}(2013)\citenamefont {Adare}
  \emph {et~al.}}]{Adare:2013piz}%
  \BibitemOpen
  \bibfield  {author} {\bibinfo {author} {\bibfnamefont {A.}~\bibnamefont
  {Adare}} \emph {et~al.} (\bibinfo {collaboration} {PHENIX Collaboration}),\
  }\bibfield  {title} {\enquote {\bibinfo {title} {{Quadrupole Anisotropy in
  Dihadron Azimuthal Correlations in Central $d$$+$Au Collisions at
  $\sqrt{s_{_{NN}}}$=200~GeV}},}\ }\href {\doibase
  10.1103/PhysRevLett.111.212301} {\bibfield  {journal} {\bibinfo  {journal}
  {Phys. Rev. Lett.}\ }\textbf {\bibinfo {volume} {111}},\ \bibinfo {pages}
  {212301} (\bibinfo {year} {2013})}\BibitemShut {NoStop}%
\bibitem [{\citenamefont {Adare}\ \emph
  {et~al.}(2015{\natexlab{a}})\citenamefont {Adare} \emph
  {et~al.}}]{Adare:2014keg}%
  \BibitemOpen
  \bibfield  {author} {\bibinfo {author} {\bibfnamefont {A.}~\bibnamefont
  {Adare}} \emph {et~al.} (\bibinfo {collaboration} {PHENIX Collaboration}),\
  }\bibfield  {title} {\enquote {\bibinfo {title} {{Measurement of long-range
  angular correlation and quadrupole anisotropy of pions and (anti)protons in
  central $d$$+$Au collisions at $\sqrt{s_{_{NN}}}$=200 GeV}},}\ }\href
  {\doibase 10.1103/PhysRevLett.114.192301} {\bibfield  {journal} {\bibinfo
  {journal} {Phys. Rev. Lett.}\ }\textbf {\bibinfo {volume} {114}},\ \bibinfo
  {pages} {192301} (\bibinfo {year} {2015}{\natexlab{a}})}\BibitemShut
  {NoStop}%
\bibitem [{\citenamefont {Adare}\ \emph
  {et~al.}(2015{\natexlab{b}})\citenamefont {Adare} \emph
  {et~al.}}]{Adare:2015ctn}%
  \BibitemOpen
  \bibfield  {author} {\bibinfo {author} {\bibfnamefont {A.}~\bibnamefont
  {Adare}} \emph {et~al.} (\bibinfo {collaboration} {PHENIX Collaboration}),\
  }\bibfield  {title} {\enquote {\bibinfo {title} {{Measurements of elliptic
  and triangular flow in high-multiplicity $^{3}$He$+$Au collisions at
  $\sqrt{s_{_{NN}}}=200$ GeV}},}\ }\href {\doibase
  10.1103/PhysRevLett.115.142301} {\bibfield  {journal} {\bibinfo  {journal}
  {Phys. Rev. Lett.}\ }\textbf {\bibinfo {volume} {115}},\ \bibinfo {pages}
  {142301} (\bibinfo {year} {2015}{\natexlab{b}})}\BibitemShut {NoStop}%
\bibitem [{\citenamefont {Aidala}\ \emph {et~al.}(2017)\citenamefont {Aidala}
  \emph {et~al.}}]{Aidala:2016vgl}%
  \BibitemOpen
  \bibfield  {author} {\bibinfo {author} {\bibfnamefont {C.}~\bibnamefont
  {Aidala}} \emph {et~al.} (\bibinfo {collaboration} {PHENIX Collaboration}),\
  }\bibfield  {title} {\enquote {\bibinfo {title} {{Measurement of long-range
  angular correlations and azimuthal anisotropies in high-multiplicity $p$$+$Au
  collisions at $\sqrt{s_{_{NN}}}=200$ GeV}},}\ }\href {\doibase
  10.1103/PhysRevC.95.034910} {\bibfield  {journal} {\bibinfo  {journal} {Phys.
  Rev. C}\ }\textbf {\bibinfo {volume} {95}},\ \bibinfo {pages} {034910}
  (\bibinfo {year} {2017})}\BibitemShut {NoStop}%
\bibitem [{\citenamefont {Khachatryan}\ \emph {et~al.}(2010)\citenamefont
  {Khachatryan} \emph {et~al.}}]{Khachatryan:2010gv}%
  \BibitemOpen
  \bibfield  {author} {\bibinfo {author} {\bibfnamefont {V.}~\bibnamefont
  {Khachatryan}} \emph {et~al.} (\bibinfo {collaboration} {CMS
  Collaboration}),\ }\bibfield  {title} {\enquote {\bibinfo {title}
  {{Observation of Long-Range Near-Side Angular Correlations in Proton-Proton
  Collisions at the LHC}},}\ }\href {\doibase 10.1007/JHEP09(2010)091}
  {\bibfield  {journal} {\bibinfo  {journal} {JHEP}\ }\textbf {\bibinfo
  {volume} {09}},\ \bibinfo {pages} {091} (\bibinfo {year} {2010})}\BibitemShut
  {NoStop}%
\bibitem [{\citenamefont {Aad}\ \emph {et~al.}(2016)\citenamefont {Aad} \emph
  {et~al.}}]{Aad:2015gqa}%
  \BibitemOpen
  \bibfield  {author} {\bibinfo {author} {\bibfnamefont {G.}~\bibnamefont
  {Aad}} \emph {et~al.} (\bibinfo {collaboration} {ATLAS Collaboration}),\
  }\bibfield  {title} {\enquote {\bibinfo {title} {{Observation of Long-Range
  Elliptic Azimuthal Anisotropies in $\sqrt{s}=$13 and 2.76~TeV $pp$ Collisions
  with the ATLAS Detector}},}\ }\href {\doibase 10.1103/PhysRevLett.116.172301}
  {\bibfield  {journal} {\bibinfo  {journal} {Phys. Rev. Lett.}\ }\textbf
  {\bibinfo {volume} {116}},\ \bibinfo {pages} {172301} (\bibinfo {year}
  {2016})}\BibitemShut {NoStop}%
\bibitem [{\citenamefont {Khachatryan}\ \emph {et~al.}(2017)\citenamefont
  {Khachatryan} \emph {et~al.}}]{Khachatryan:2016txc}%
  \BibitemOpen
  \bibfield  {author} {\bibinfo {author} {\bibfnamefont {V.}~\bibnamefont
  {Khachatryan}} \emph {et~al.} (\bibinfo {collaboration} {CMS
  Collaboration}),\ }\bibfield  {title} {\enquote {\bibinfo {title} {{Evidence
  for collectivity in $pp$ collisions at the LHC}},}\ }\href {\doibase
  10.1016/j.physletb.2016.12.009} {\bibfield  {journal} {\bibinfo  {journal}
  {Phys. Lett. B}\ }\textbf {\bibinfo {volume} {765}},\ \bibinfo {pages} {193}
  (\bibinfo {year} {2017})}\BibitemShut {NoStop}%
\bibitem [{\citenamefont {Aad}\ \emph {et~al.}(2013{\natexlab{b}})\citenamefont
  {Aad} \emph {et~al.}}]{Aad:2013fja}%
  \BibitemOpen
  \bibfield  {author} {\bibinfo {author} {\bibfnamefont {G.}~\bibnamefont
  {Aad}} \emph {et~al.} (\bibinfo {collaboration} {ATLAS Collaboration}),\
  }\bibfield  {title} {\enquote {\bibinfo {title} {{Measurement with the ATLAS
  detector of multi-particle azimuthal correlations in p+Pb collisions at
  $\sqrt{s_{NN}}$=5.02~TeV}},}\ }\href {\doibase
  10.1016/j.physletb.2013.06.057} {\bibfield  {journal} {\bibinfo  {journal}
  {Phys. Lett. B}\ }\textbf {\bibinfo {volume} {725}},\ \bibinfo {pages} {60}
  (\bibinfo {year} {2013}{\natexlab{b}})}\BibitemShut {NoStop}%
\bibitem [{\citenamefont {Chatrchyan}\ \emph
  {et~al.}(2013{\natexlab{b}})\citenamefont {Chatrchyan} \emph
  {et~al.}}]{Chatrchyan:2013nka}%
  \BibitemOpen
  \bibfield  {author} {\bibinfo {author} {\bibfnamefont {S.}~\bibnamefont
  {Chatrchyan}} \emph {et~al.} (\bibinfo {collaboration} {CMS Collaboration}),\
  }\bibfield  {title} {\enquote {\bibinfo {title} {{Multiplicity and transverse
  momentum dependence of two- and four-particle correlations in pPb and PbPb
  collisions}},}\ }\href {\doibase 10.1016/j.physletb.2013.06.028} {\bibfield
  {journal} {\bibinfo  {journal} {Phys. Lett. B}\ }\textbf {\bibinfo {volume}
  {724}},\ \bibinfo {pages} {213} (\bibinfo {year}
  {2013}{\natexlab{b}})}\BibitemShut {NoStop}%
\bibitem [{\citenamefont {Abelev}\ \emph {et~al.}(2014)\citenamefont {Abelev}
  \emph {et~al.}}]{Abelev:2014mda}%
  \BibitemOpen
  \bibfield  {author} {\bibinfo {author} {\bibfnamefont {B.~B.}\ \bibnamefont
  {Abelev}} \emph {et~al.} (\bibinfo {collaboration} {ALICE Collaboration}),\
  }\bibfield  {title} {\enquote {\bibinfo {title} {{Multiparticle azimuthal
  correlations in p-Pb and Pb-Pb collisions at the CERN Large Hadron
  Collider}},}\ }\href {\doibase 10.1103/PhysRevC.90.054901} {\bibfield
  {journal} {\bibinfo  {journal} {Phys. Rev. C}\ }\textbf {\bibinfo {volume}
  {90}},\ \bibinfo {pages} {054901} (\bibinfo {year} {2014})}\BibitemShut
  {NoStop}%
\bibitem [{\citenamefont {Khachatryan}\ \emph {et~al.}(2015)\citenamefont
  {Khachatryan} \emph {et~al.}}]{Khachatryan:2015waa}%
  \BibitemOpen
  \bibfield  {author} {\bibinfo {author} {\bibfnamefont {V.}~\bibnamefont
  {Khachatryan}} \emph {et~al.} (\bibinfo {collaboration} {CMS
  Collaboration}),\ }\bibfield  {title} {\enquote {\bibinfo {title} {{Evidence
  for Collective Multiparticle Correlations in p-Pb Collisions}},}\ }\href
  {\doibase 10.1103/PhysRevLett.115.012301} {\bibfield  {journal} {\bibinfo
  {journal} {Phys. Rev. Lett.}\ }\textbf {\bibinfo {volume} {115}},\ \bibinfo
  {pages} {012301} (\bibinfo {year} {2015})}\BibitemShut {NoStop}%
\bibitem [{\citenamefont {Loizides}(2016)}]{Loizides:2016tew}%
  \BibitemOpen
  \bibfield  {author} {\bibinfo {author} {\bibfnamefont {C.}~\bibnamefont
  {Loizides}},\ }\bibfield  {title} {\enquote {\bibinfo {title} {{Experimental
  overview on small collision systems at the LHC}},}\ }\bibfield  {booktitle}
  {\emph {\bibinfo {booktitle} {{Proceedings, 25th International Conference on
  Ultra-Relativistic Nucleus-Nucleus Collisions (Quark Matter 2015): Kobe,
  Japan, September 27-October 3, 2015}}},\ }\href {\doibase
  10.1016/j.nuclphysa.2016.04.022} {\bibfield  {journal} {\bibinfo  {journal}
  {Nucl. Phys. A}\ }\textbf {\bibinfo {volume} {956}},\ \bibinfo {pages} {200}
  (\bibinfo {year} {2016})}\BibitemShut {NoStop}%
\bibitem [{\citenamefont {Dusling}\ \emph {et~al.}()\citenamefont {Dusling},
  \citenamefont {Mace},\ and\ \citenamefont {Venugopalan}}]{Dusling:2017aot}%
  \BibitemOpen
  \bibfield  {author} {\bibinfo {author} {\bibfnamefont {K.}~\bibnamefont
  {Dusling}}, \bibinfo {author} {\bibfnamefont {M.}~\bibnamefont {Mace}}, \
  and\ \bibinfo {author} {\bibfnamefont {R.}~\bibnamefont {Venugopalan}},\
  }\href@noop {} {\enquote {\bibinfo {title} {{Parton model description of
  multiparticle azimuthal correlations in $pA$ collisions}},}\ }\bibinfo {note}
  {ArXiv:1706.06260}\BibitemShut {NoStop}%
\bibitem [{\citenamefont {Voloshin}\ and\ \citenamefont
  {Zhang}(1996)}]{Voloshin:1994mz}%
  \BibitemOpen
  \bibfield  {author} {\bibinfo {author} {\bibfnamefont {S.~A.}\ \bibnamefont
  {Voloshin}}\ and\ \bibinfo {author} {\bibfnamefont {Y.}~\bibnamefont
  {Zhang}},\ }\bibfield  {title} {\enquote {\bibinfo {title} {{Flow study in
  relativistic nuclear collisions by Fourier expansion of Azimuthal particle
  distributions}},}\ }\href {\doibase 10.1007/s002880050141} {\bibfield
  {journal} {\bibinfo  {journal} {Z. Phys. C}\ }\textbf {\bibinfo {volume}
  {70}},\ \bibinfo {pages} {665} (\bibinfo {year} {1996})}\BibitemShut
  {NoStop}%
\bibitem [{\citenamefont {Bilandzic}\ \emph {et~al.}(2011)\citenamefont
  {Bilandzic}, \citenamefont {Snellings},\ and\ \citenamefont
  {Voloshin}}]{Bilandzic:2010jr}%
  \BibitemOpen
  \bibfield  {author} {\bibinfo {author} {\bibfnamefont {A.}~\bibnamefont
  {Bilandzic}}, \bibinfo {author} {\bibfnamefont {R.}~\bibnamefont
  {Snellings}}, \ and\ \bibinfo {author} {\bibfnamefont {S.}~\bibnamefont
  {Voloshin}},\ }\bibfield  {title} {\enquote {\bibinfo {title} {{Flow analysis
  with cumulants: Direct calculations}},}\ }\href {\doibase
  10.1103/PhysRevC.83.044913} {\bibfield  {journal} {\bibinfo  {journal} {Phys.
  Rev. C}\ }\textbf {\bibinfo {volume} {83}},\ \bibinfo {pages} {044913}
  (\bibinfo {year} {2011})}\BibitemShut {NoStop}%
\bibitem [{\citenamefont {Ollitrault}\ \emph {et~al.}(2009)\citenamefont
  {Ollitrault}, \citenamefont {Poskanzer},\ and\ \citenamefont
  {Voloshin}}]{Ollitrault:2009ie}%
  \BibitemOpen
  \bibfield  {author} {\bibinfo {author} {\bibfnamefont {J.-Y.}\ \bibnamefont
  {Ollitrault}}, \bibinfo {author} {\bibfnamefont {A.~M.}\ \bibnamefont
  {Poskanzer}}, \ and\ \bibinfo {author} {\bibfnamefont {S.~A.}\ \bibnamefont
  {Voloshin}},\ }\bibfield  {title} {\enquote {\bibinfo {title} {{Effect of
  flow fluctuations and nonflow on elliptic flow methods}},}\ }\href {\doibase
  10.1103/PhysRevC.80.014904} {\bibfield  {journal} {\bibinfo  {journal} {Phys.
  Rev. C}\ }\textbf {\bibinfo {volume} {80}},\ \bibinfo {pages} {014904}
  (\bibinfo {year} {2009})}\BibitemShut {NoStop}%
\bibitem [{\citenamefont {Adcox}\ \emph {et~al.}(2003)\citenamefont {Adcox}
  \emph {et~al.}}]{Adcox:2003zm}%
  \BibitemOpen
  \bibfield  {author} {\bibinfo {author} {\bibfnamefont {K.}~\bibnamefont
  {Adcox}} \emph {et~al.} (\bibinfo {collaboration} {PHENIX Collaboration}),\
  }\bibfield  {title} {\enquote {\bibinfo {title} {{PHENIX detector
  overview}},}\ }\href {\doibase 10.1016/S0168-9002(02)01950-2} {\bibfield
  {journal} {\bibinfo  {journal} {Nucl. Instrum. Methods Phys. Res., Sec. A}\
  }\textbf {\bibinfo {volume} {499}},\ \bibinfo {pages} {469} (\bibinfo {year}
  {2003})}\BibitemShut {NoStop}%
\bibitem [{\citenamefont {Ikematsu}\ \emph {et~al.}(1998)\citenamefont
  {Ikematsu} \emph {et~al.}}]{Ikematsu:1998fm}%
  \BibitemOpen
  \bibfield  {author} {\bibinfo {author} {\bibfnamefont {K.}~\bibnamefont
  {Ikematsu}} \emph {et~al.},\ }\bibfield  {title} {\enquote {\bibinfo {title}
  {{A Start- timing detector for the collider experiment PHENIX at
  RHIC-BNL}},}\ }\href {\doibase 10.1016/S0168-9002(98)00307-6} {\bibfield
  {journal} {\bibinfo  {journal} {Nucl. Instrum. Methods Phys. Res., Sec. A}\
  }\textbf {\bibinfo {volume} {411}},\ \bibinfo {pages} {238} (\bibinfo {year}
  {1998})}\BibitemShut {NoStop}%
\bibitem [{\citenamefont {Aidala}\ \emph {et~al.}(2014)\citenamefont {Aidala}
  \emph {et~al.}}]{Aidala:2013vna}%
  \BibitemOpen
  \bibfield  {author} {\bibinfo {author} {\bibfnamefont {C.}~\bibnamefont
  {Aidala}} \emph {et~al.},\ }\bibfield  {title} {\enquote {\bibinfo {title}
  {{The PHENIX Forward Silicon Vertex Detector}},}\ }\href {\doibase
  10.1016/j.nima.2014.04.017} {\bibfield  {journal} {\bibinfo  {journal} {Nucl.
  Instrum. Methods Phys. Res., Sec. A}\ }\textbf {\bibinfo {volume} {755}},\
  \bibinfo {pages} {441} (\bibinfo {year} {2014})}\BibitemShut {NoStop}%
\bibitem [{\citenamefont {Poskanzer}\ and\ \citenamefont
  {Voloshin}(1998)}]{Poskanzer:1998yz}%
  \BibitemOpen
  \bibfield  {author} {\bibinfo {author} {\bibfnamefont {A.~M.}\ \bibnamefont
  {Poskanzer}}\ and\ \bibinfo {author} {\bibfnamefont {S.~A.}\ \bibnamefont
  {Voloshin}},\ }\bibfield  {title} {\enquote {\bibinfo {title} {{Methods for
  analyzing anisotropic flow in relativistic nuclear collisions}},}\ }\href
  {\doibase 10.1103/PhysRevC.58.1671} {\bibfield  {journal} {\bibinfo
  {journal} {Phys. Rev. C}\ }\textbf {\bibinfo {volume} {58}},\ \bibinfo
  {pages} {1671} (\bibinfo {year} {1998})}\BibitemShut {NoStop}%
\bibitem [{\citenamefont {Lin}\ \emph {et~al.}(2005)\citenamefont {Lin},
  \citenamefont {Ko}, \citenamefont {Li}, \citenamefont {Zhang},\ and\
  \citenamefont {Pal}}]{Lin:2004en}%
  \BibitemOpen
  \bibfield  {author} {\bibinfo {author} {\bibfnamefont {Z.-W.}\ \bibnamefont
  {Lin}}, \bibinfo {author} {\bibfnamefont {C.~M.}\ \bibnamefont {Ko}},
  \bibinfo {author} {\bibfnamefont {B.-A.}\ \bibnamefont {Li}}, \bibinfo
  {author} {\bibfnamefont {B.}~\bibnamefont {Zhang}}, \ and\ \bibinfo {author}
  {\bibfnamefont {S.}~\bibnamefont {Pal}},\ }\bibfield  {title} {\enquote
  {\bibinfo {title} {{A Multi-phase transport model for relativistic heavy ion
  collisions}},}\ }\href {\doibase 10.1103/PhysRevC.72.064901} {\bibfield
  {journal} {\bibinfo  {journal} {Phys. Rev. C}\ }\textbf {\bibinfo {volume}
  {72}},\ \bibinfo {pages} {064901} (\bibinfo {year} {2005})}\BibitemShut
  {NoStop}%
\bibitem [{\citenamefont {Bzdak}\ and\ \citenamefont
  {Ma}(2014)}]{Bzdak:2014dia}%
  \BibitemOpen
  \bibfield  {author} {\bibinfo {author} {\bibfnamefont {A.}~\bibnamefont
  {Bzdak}}\ and\ \bibinfo {author} {\bibfnamefont {G.-L.}\ \bibnamefont {Ma}},\
  }\bibfield  {title} {\enquote {\bibinfo {title} {{Elliptic and triangular
  flow in $p$$+$Pb and peripheral Pb$+$Pb collisions from parton
  scatterings}},}\ }\href {\doibase 10.1103/PhysRevLett.113.252301} {\bibfield
  {journal} {\bibinfo  {journal} {Phys. Rev. Lett.}\ }\textbf {\bibinfo
  {volume} {113}},\ \bibinfo {pages} {252301} (\bibinfo {year}
  {2014})}\BibitemShut {NoStop}%
\bibitem [{\citenamefont {Orjuela~Koop}\ \emph {et~al.}(2015)\citenamefont
  {Orjuela~Koop}, \citenamefont {Adare}, \citenamefont {McGlinchey},\ and\
  \citenamefont {Nagle}}]{Koop:2015wea}%
  \BibitemOpen
  \bibfield  {author} {\bibinfo {author} {\bibfnamefont {J.~D.}\ \bibnamefont
  {Orjuela~Koop}}, \bibinfo {author} {\bibfnamefont {A.}~\bibnamefont {Adare}},
  \bibinfo {author} {\bibfnamefont {D.}~\bibnamefont {McGlinchey}}, \ and\
  \bibinfo {author} {\bibfnamefont {J.~L.}\ \bibnamefont {Nagle}},\ }\bibfield
  {title} {\enquote {\bibinfo {title} {{Azimuthal anisotropy relative to the
  participant plane from a multiphase transport model in central $p$$+$Au,
  $d$$+$Au, and $^{3}$He$+$Au collisions at $\sqrt{s_{NN}}=200$~GeV}},}\ }\href
  {\doibase 10.1103/PhysRevC.92.054903} {\bibfield  {journal} {\bibinfo
  {journal} {Phys. Rev. C}\ }\textbf {\bibinfo {volume} {92}},\ \bibinfo
  {pages} {054903} (\bibinfo {year} {2015})}\BibitemShut {NoStop}%
\bibitem [{\citenamefont {Orjuela~Koop}\ \emph {et~al.}(2016)\citenamefont
  {Orjuela~Koop}, \citenamefont {Belmont}, \citenamefont {Yin},\ and\
  \citenamefont {Nagle}}]{Koop:2015trj}%
  \BibitemOpen
  \bibfield  {author} {\bibinfo {author} {\bibfnamefont {J.~D.}\ \bibnamefont
  {Orjuela~Koop}}, \bibinfo {author} {\bibfnamefont {R.}~\bibnamefont
  {Belmont}}, \bibinfo {author} {\bibfnamefont {P.}~\bibnamefont {Yin}}, \ and\
  \bibinfo {author} {\bibfnamefont {J.~L.}\ \bibnamefont {Nagle}},\ }\bibfield
  {title} {\enquote {\bibinfo {title} {{Exploring the Beam Energy Dependence of
  Flow-Like Signatures in Small System $d$$+$Au Collisions}},}\ }\href
  {\doibase 10.1103/PhysRevC.93.044910} {\bibfield  {journal} {\bibinfo
  {journal} {Phys. Rev. C}\ }\textbf {\bibinfo {volume} {93}},\ \bibinfo
  {pages} {044910} (\bibinfo {year} {2016})}\BibitemShut {NoStop}%
\bibitem [{\citenamefont {Loizides}\ \emph {et~al.}(2015)\citenamefont
  {Loizides}, \citenamefont {Nagle},\ and\ \citenamefont
  {Steinberg}}]{Loizides:2014vua}%
  \BibitemOpen
  \bibfield  {author} {\bibinfo {author} {\bibfnamefont {C.}~\bibnamefont
  {Loizides}}, \bibinfo {author} {\bibfnamefont {J.}~\bibnamefont {Nagle}}, \
  and\ \bibinfo {author} {\bibfnamefont {P.}~\bibnamefont {Steinberg}},\
  }\bibfield  {title} {\enquote {\bibinfo {title} {{Improved version of the
  PHOBOS Glauber Monte Carlo}},}\ }\href {\doibase 10.1016/j.softx.2015.05.001}
  {\bibfield  {journal} {\bibinfo  {journal} {SoftwareX}\ }\textbf {\bibinfo
  {volume} {1}},\ \bibinfo {pages} {13} (\bibinfo {year} {2015})}\BibitemShut
  {NoStop}%
\bibitem [{\citenamefont {Petersen}\ \emph {et~al.}(2011)\citenamefont
  {Petersen}, \citenamefont {Bhattacharya}, \citenamefont {Bass},\ and\
  \citenamefont {Greiner}}]{Petersen:2011fp}%
  \BibitemOpen
  \bibfield  {author} {\bibinfo {author} {\bibfnamefont {H.}~\bibnamefont
  {Petersen}}, \bibinfo {author} {\bibfnamefont {V.}~\bibnamefont
  {Bhattacharya}}, \bibinfo {author} {\bibfnamefont {S.~A.}\ \bibnamefont
  {Bass}}, \ and\ \bibinfo {author} {\bibfnamefont {C.}~\bibnamefont
  {Greiner}},\ }\bibfield  {title} {\enquote {\bibinfo {title} {{Longitudinal
  correlation of the triangular flow event plane in a hybrid approach with
  hadron and parton cascade initial conditions}},}\ }\href {\doibase
  10.1103/PhysRevC.84.054908} {\bibfield  {journal} {\bibinfo  {journal} {Phys.
  Rev. C}\ }\textbf {\bibinfo {volume} {84}},\ \bibinfo {pages} {054908}
  (\bibinfo {year} {2011})}\BibitemShut {NoStop}%
\bibitem [{\citenamefont {Xiao}\ \emph {et~al.}(2013)\citenamefont {Xiao},
  \citenamefont {Liu},\ and\ \citenamefont {Wang}}]{Xiao:2012uw}%
  \BibitemOpen
  \bibfield  {author} {\bibinfo {author} {\bibfnamefont {K.}~\bibnamefont
  {Xiao}}, \bibinfo {author} {\bibfnamefont {F.}~\bibnamefont {Liu}}, \ and\
  \bibinfo {author} {\bibfnamefont {F.}~\bibnamefont {Wang}},\ }\bibfield
  {title} {\enquote {\bibinfo {title} {{Event-plane decorrelation over
  pseudorapidity and its effect on azimuthal anisotropy measurements in
  relativistic heavy-ion collisions}},}\ }\href {\doibase
  10.1103/PhysRevC.87.011901} {\bibfield  {journal} {\bibinfo  {journal} {Phys.
  Rev. C}\ }\textbf {\bibinfo {volume} {87}},\ \bibinfo {pages} {011901}
  (\bibinfo {year} {2013})}\BibitemShut {NoStop}%
\bibitem [{\citenamefont {Ollitrault}(1992)}]{PhysRevD.46.229}%
  \BibitemOpen
  \bibfield  {author} {\bibinfo {author} {\bibfnamefont {J.-Y.}\ \bibnamefont
  {Ollitrault}},\ }\bibfield  {title} {\enquote {\bibinfo {title} {Anisotropy
  as a signature of transverse collective flow},}\ }\href {\doibase
  10.1103/PhysRevD.46.229} {\bibfield  {journal} {\bibinfo  {journal} {Phys.
  Rev. D}\ }\textbf {\bibinfo {volume} {46}},\ \bibinfo {pages} {229} (\bibinfo
  {year} {1992})}\BibitemShut {NoStop}%
\end{thebibliography}

%
 
\end{document}